\documentclass[a4paper,referee]{aa}
\usepackage{natbib}
\usepackage{rotating}

\usepackage{longtable}
\usepackage{lscape}

\begin{document}

\title{Spectroscopic observations of blue stars with infrared excesses in NGC 6611}

\author{R. Bonito\inst{1, 2} \and L. Prisinzano\inst{2} \and M. G. Guarcello\inst{3}
\and G. Micela\inst{2}}

\offprints{R. Bonito\\ \email{sbonito@astropa.unipa.it}}

\institute{Dipartimento di Fisica e Chimica, Universit\`a di Palermo, P.zza del Parlamento 1, 90134
Palermo, Italy
\and 
INAF -- Osservatorio Astronomico di Palermo, P.zza del Parlamento 1,
90134 Palermo, Italy 
\and
Smithsonian Astrophysical Observatory - SAO} 

\date{Received, accepted}

\authorrunning{}
\titlerunning{BWE stars in NGC 6611}

\abstract
%Context. 
{The young open cluster NGC $6611$ includes among its candidate members a class of peculiar objects with interesting properties: blue stars with infrared IR excesses.
These stars show excesses in IR bands, signature of the presence of a circumstellar disk, but optical colors typical of older field stars. In order to confirm their membership to the cluster, it is therefore important to use new spectroscopic observations, together with previous photometric data. }  
%Aims. 
{We aim at confirming the membership of these objects and at investigating their physical properties to verify whether the observed colors are intrinsic or altered by the presence of the disk or by the accretion processes.  }
%Methods. 
{We analyze the intermediate resolution spectroscopic data obtained for a subsample of blue stars in NGC $6611$ with FLAMES. In particular, we focus on the study of: 1) the profile of the $H\alpha$ emission line, to select stars with accretion and outflow activity; 2) the $Li$ absorption line, used as a youth indicator; 3) the radial velocity. }
%Results. 
{Using the spectroscopic analysis, it has been possible to investigate the presence of the $Li$ absorption line as well as to discriminate between stars with inert or active disk. In particular, from the analysis of the $H\alpha$ emission line we were able to infer the activity due to the accretion and outflow processes and the variability of the emission. We also investigated the binarity of the blue stars and their membership to NGC $6611$. }
%Conclusion. 
{From our spectroscopic analysis, we conclude that half of the sample of blue stars ($10/20$) are confirmed members of NGC $6611$ (with $6$ more stars that could also be possible members).
In conclusion, our results indicate that members of young clusters can be found also in an anomalous region of the color-magnitude diagram, i.e. outside of the pre main sequence locus were most of the cluster members lie.}

\keywords{stars: formation - stars: pre-main sequence - accretion, accretion disks - individual: NGC 6611}

\maketitle

\section{Introduction}
\label{Introduction}

The evolution of stars in the pre main sequence (PMS) phase can be investigated by studying the properties of young stellar objects (YSOs) in star forming regions. 
Spectral analysis of young clusters can shed a light on the accretion properties of YSOs, in particular in the age range $1 - 10$ Myr, i.e. when the YSOs lose their disks and the accretion process stops (\citealt{hcg98}).
Disk evolution and properties are interesting also in the wider context of planetary formation.

In this paper we focus on the NGC $6611$ cluster. This cluster and its parental cloud M $16$ (the Eagle Nebula) have been deeply investigated in the past and the main parameters have been derived in detail (\citealt{gpm07,gdm10}). 
The authors estimated a distance of $1750$ pc, a median age of $\sim 1$ Myr, and an average disk frequency of $\approx 36\%$. This value however increases far from the OB stars.
Previous studies of the central region of NGC $6611$ claim large age spread, $< 1 - 3$ Myrs (\citealt{irw07}), and different values for the disk frequency, $58\%$ (\citealt{ojv05}).

In this paper, we study the nature of a group of stars, called Blue With infrared Excesses (BWE), found in NGC 6611 by \citet{gdm10}. Young stars associated with NGC $6611$ were selected by those authors (see Sect. \ref{Observations}) by analyzing their photometric data, by checking their infrared (IR) excesses, which are a signature of the presence of a circumstellar disk, and their detection in the X-ray band. The
stars with infrared excesses are consistent with ages ``apparently" older than the cluster age, i.e. with colors bluer than those predicted by the 5 Myrs isochrone (chosen to be conservative) of \citet{sdf00} at the distance and extinction of the cluster. Our aim is to understand if these stars are members of NGC $6611$ and if their ``anomalous" position on the color-magnitude diagram (CMD) is due to the presence of the circumstellar disk and/or to the accretion process or if their colors are purely photospheric, leading to the conclusion that these stars are older than the other stars of the cluster.

The use of spectroscopic data is necessary to distinguish between classical T Tauri stars (CTTSs) and weak line T Tauri stars (WTTSs). In fact, using only photometric data it is not possible to discriminate between these two classes, since we know that there are stars with a disk which do not show accretion activity. Indeed, the fraction of disks without accretion activity grows with age (\citealt{sfr08}). 
This implies different time-scales for the two effects, the lifetime of the 
circumstellar disk and the accretion timescales.
In particular, the spectroscopic analysis allows us to study the $H\alpha$ emission line, important since we can infer from its profile the presence of accretion and outflow. In fact, CTTSs spectra show broad $H\alpha$ emission lines, due to the accretion and outflow processes, while WTTSs do not show any feature of accretion neither of outflow. 
Furthermore, spectra containing the absorption $Li$ line allow us to discriminate between young and old stars, since the presence of this line is an indicator of youth. 
In fact, during the PMS, in stars with $M < 1.2 M_{\odot}$ (where the convective region is more extended) the Li in the stellar surface reaches the internal region where it is destroyed. The process is very efficient in particular in totally convective stars with mass between $0.2 - 0.5 M_{\odot}$ where the Li is depleted in $5 - 10$ Myr (\citealt{bod65}; \citealt{sdf00}).

In order to understand the nature of the BWE stars, several interpretations have been proposed (see \citealt{gdm10} for details). 
One hypothesis is that the V band can be altered by accretion activity. In particular, the photospheric colors of the stars can be modified by the veiling effect (as in the Taurus-Auriga complex discussed by \citealt{hk90}).
Furthermore, the scattering and the obscuration due to the disk can affect the optical bands more than the IR bands, leading to a V-I bluer than expected. In fact, scattering due to small dust grains in circumstellar disks can cause an increase of the optical flux at shorter wavelength (\citealt{tbe01}), while the obscuration due to a high inclination angle of the disk can reduce the photospheric flux, absorbing part of the stellar radiation, without altering the ($V-I$) colors. Therefore the scattering can make the emission bluer, while the obscuration makes the optical emission fainter. 

It is worth noting that a mismatch between the optical and the IR catalogs could misinterpret MS optical foreground stars as BWE candidate members. This can happen since the most probable mismatch occurs between a foreground optical source and a background infrared source. The combined spectral energy distribution (SED) can resemble that of a YSO with large infrared excesses, that will be easily detected by the selection criteria adopted in \citet{gpm07} which combine and compare optical and infrared colors.
Note that as all the optical data have been collected during the same night (\citealt{gpm07}), no photometric variability can be considered as a cause for the observed blue V-I colors.

If the BWE stars were old cluster members, this should be evidence of age spread and this is crucial to constrain the theoretical models of cluster formation (e.g.  \citealt{sal87}, \citealt{tm04}, \citealt{bph07}).

The consequences of our study are of general interest: BWE stars are typically excluded from cluster memberships based on photometric data. If a significant fraction of the BWE stars analyzed here are spectroscopically confirmed members of NGC $6611$, this will imply that not including them among candidate members creates a serious completeness issue, with consequences on studies, for instance, on the IMF and cluster dynamics.

We note that stars with properties similar to those of BWE stars have been observed not only in NGC $6611$ but also in other clusters (as in the ONC, \citealt{hil97}; see also Sect. \ref{membership} for more examples), therefore the analysis discussed here has a wide application.

\section{Target selection and observations}
\label{Observations}

In the context of the ESO program Obs. ID: 083.C-0837 (P.I.: Guarcello), we observed a sample of $194$ candidate members of the cluster NGC $6611$, selected from the \citet{gdm10} catalog, with the aim of deriving their spectroscopic properties. 

Disk bearing cluster members have been selected from \citet{gpm07,gmd09,gdm10} adopting two criteria: the IRAC $[3.6]-[4.5]$ vs. $[5.8]-[8.0]$ diagram and a set of reddening-free color indices.  In the IRAC color-color diagram the locus of normal-color stars, reddened photospheres, and disk-bearing objects can be easily separated, resulting in a robust selection suffering a small contamination (\citealt{acd04}). The set of reddening free color indices used for the selection is defined in \citet{dpm06} and \citet{gpm07,gmd09}. 
Since a detailed explanation of the properties of these indices is beyond the scope of this paper, we just summarize their main characteristics. The reader is referred to these papers for more details. Each of these indices is defined in order to compare an optical color, representative of the photospheric emission, with an infrared color. Normal stars have indices with positive values even at large extinctions, while the indices become more negative when stars have infrared excesses, being possible to build diagrams where the locus of the reddened stars is well distinguished from that of the disk-bearing objects. Besides, the indices are particularly effective (\citealt{gdm10}) in selecting stars whose optical colors are affected by accretion or presence of light scattered into the line of sigh by the disk itself.

While most of these sources show ages in agreement with the cluster age ($< 5$ Myr), a sub-sample of $90$ stars with disk show optical colors typical of objects older than the remainder cluster members (the BWE stars). In this paper we analyze $20$ of these stars suitable for FLAMES observations. Table \ref{fotometria} shows the list of the $20$ BWE stars analyzed here with the related B, V, and I magnitudes.

\begin{table}
\begin{center}
\caption{List of the $20$ BWE stars observed with FLAMES and their B, V, and I magnitudes.
\label{fotometria}}
\begin{tabular}{crrrrrrrrrrr}
%\tableline\tableline
\hline
\hline
BWE ID & B & V & I \\
\hline
10151 & $18.083$ & $17.091$ & $15.558$ \\
11447 & $18.171$ & $16.949$ & $15.582$ \\
11532 & $18.59$  & $17.349$ & $15.952$ \\
11598 & $18.268$ & $17.186$ & $15.899$ \\
11751 & $16.598$ & $15.733$ & $14.502$ \\
1211  & $19.166$ & $17.69$  & $16.151$ \\
12168 & $18.936$ & $17.693$ & $16.187$ \\
1233  & $18.93$  & $17.626$ & $15.978$ \\
12631 & $18.61$  & $17.358$ & $15.922$ \\
1320  & $16.898$ & $15.778$ & $14.576$ \\
13310 & $19.409$ & $17.971$ & $16.195$ \\
1336  & $17.736$ & $16.631$ & $15.203$ \\
13407 & $19.251$ & $17.749$ & $15.965$ \\
1455  & $19.45$  & $17.874$ & $16.292$ \\
15584 & $18.075$ & $16.595$ & $14.912$ \\
15805 & $19.147$ & $17.509$ & $15.709$ \\
15806 & $20.462$ & $18.609$ & $16.422$ \\
2062  & $17.222$ & $16.321$ & $14.982$ \\
2594  & $17.659$ & $16.452$ & $15.119$ \\
4112  & $19.18$  & $17.674$ & $15.883$ \\
\hline
\end{tabular}
\end{center}
\end{table}

\begin{figure}[!t]
\centerline{\includegraphics[angle=0,width=10cm]{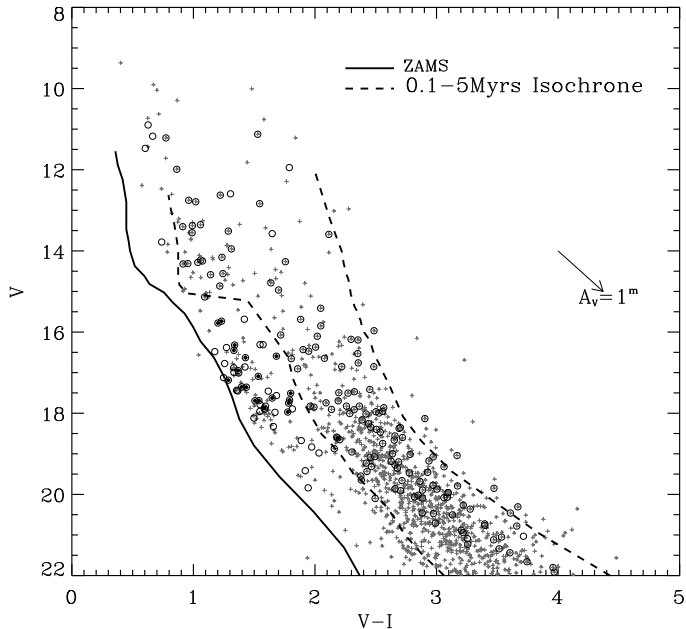}}
\caption{$V$ vs. $V - I$ color-magnitude diagram of NGC $6611$ members from the \citet{gdm10} catalog (crosses). The $20$ BWE stars analyzed here are marked with black filled circles, while the empty circles are all the targets observed with FLAMES. 
The solid and dashed lines are the ZAMS and the isochrone at  $0.1 - 5$ Myr, respectively (\citealt{sdf00}). The reddening vector is also shown (\citealt{mc96}). Targets have been selected imposing $V-I < 0.15$ mag.}
\label{VVI-BWE}
\end{figure}

Figure \ref{VVI-BWE} shows the $V$ vs. $V - I$ color-magnitude diagram of the cluster members from the \citet{gdm10} catalog (crosses), with all the FLAMES targets indicated with empty circles, and the $20$ BWE stars analyzed here marked with black filled circles (the error associated to $V-I$ requested for the selection of the targets is $<0.15$ mag).

The spectra analyzed in this program have been obtained with the GIRAFFE/FLAMES@VLT instrument operating in MEDUSA mode using the HR15N setup. 
The spectra have a resolution $R \approx 17000$ and a signal to noise ratio $S/N \approx 20 - 90$. 
The spectral range of this setup includes the $Li~ 6707.8$~\AA~absorption line as well as the $H\alpha~ 6563$~\AA~emission line. 

The details on the FLAMES observations are summarized in Table \ref{obs-log}.

\begin{table*}
\begin{center}
\caption{Observations log.
\label{obs-log}}
\begin{tabular}{crrrrrrrrrrr}
\hline
\hline
Config. ID & Setup & $t_{exp} (s)$ & Run dates & OB name\\
\hline
354292 & HR15N & 2760.0167 & 2009-05-22 & OBA \\
354294 & HR15N & 2760.0153 & 2009-06-19 & OBB \\
354296 & HR15N & 2760.0152 & 2009-06-19 & OBC \\

354298 & HR15N & 2760.0146 & 2009-06-27 & OBA-2 \\
354300 & HR15N & 2760.0154 & 2009-06-27 & OBB-2 \\
354302 & HR15N & 2760.0163 & 2009-06-27 & OBC-2 \\
\hline
\end{tabular}
\end{center}
\end{table*}

\section{Results}
\label{Results}

It is important to use independent criteria to confirm the origin of the BWE stars and their membership to NGC $6611$.
Here we focus mainly on the analysis of the $Li$ and $H\alpha$ lines, the radial velocity, $RV$, and $vsin(i)$ measurements, using also the X-ray detection previously obtained. 

\subsection{Lithium equivalent width}
\label{Li}

The presence of a strong absorption $Li$ line is commonly used as a membership criterion as it is indicative of youth of stars. 
The evidence of a strong $Li$ line is a membership criterion also for WTTSs (therefore it is a complementary criterion with respect to the study of the $H\alpha$ profile discussed in Sect. \ref{Ha}).

In order to subtract the background contribution due to the sky, we computed the median of several ($\approx 16 - 20$) spectra of the sky for each observation during the same night (same OB) and subtract it to the spectrum of the BWE analyzed (following the method of \citealt{jo05}). 
After the subtraction of the median sky value, we have corrected for the $RV$ measurements (presented in Sect. \ref{RV and vsini}).
Then we combined the different OBs by summing the spectra, in order to improve the S/N.
To derive the the equivalent width, $EW$, of the $Li$ line of the BWE stars,  $EW(Li)$, we first normalized the spectrum of each BWE star to the continuum using the CONTINUUM task of IRAF in a small ($10$ \AA) region centered on the $Li$ absorption line at $6707.8$ \AA, using a second order Legendre function and a variable residual rejection limit. Then, we measured the $EW(Li)$ by using the SPLOT task of IRAF. In particular, we fit the line with a Voigt profile if its $EW$ is $> 100$ m\AA~ and with a Gaussian if its $EW$ is $<100$  m\AA. 
The contribution due to the nearby Fe I line at 6707.441 \AA~ blended with the $Li$ line is negligible with respect to the $EW(Li)$. In fact, following \citet{sjb93}, we derive that $EW(Fe)$ ranges between $5$ and $70$ m\AA, while the $EW(Li)$ measured for the BWE stars showing significant $Li$ absorption line is $> 180$ m\AA. Therefore, we do not correct the $EW(Li)$ for the contribution of the iron line.
See Table \ref{res} for the $EW(Li)$ values derived for BWE stars with $EW(Li) > 100$ m\AA.
The strongest $Li$ lines detected have $EW$ varying between $\approx190 - 450$ m\AA, values consistent with other very young clusters (e.g. \citealt{pdm07}).

\begin{figure}[!t]
\centerline{\includegraphics[angle=0,width=8cm]{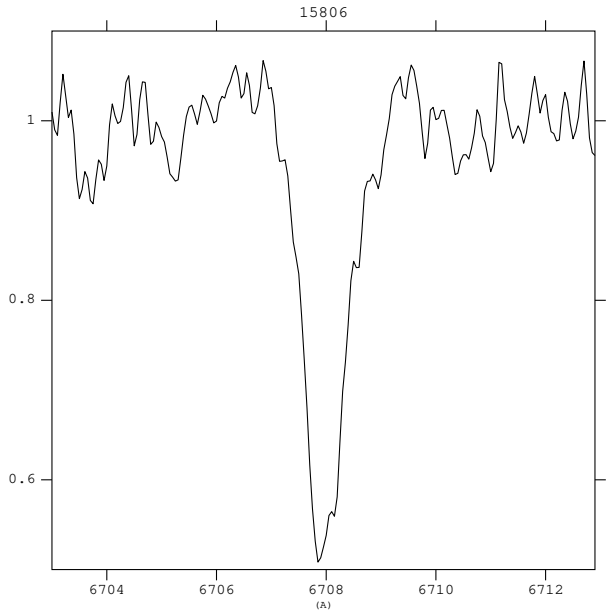}}
\caption{The strong $Li$ line in the BWE star $15806$. We have subtracted the median sky value, we have corrected for the $RV$ measurements, and we combined the different OBs by summing the spectra.}
\label{EWLi-15806}
\end{figure}

An example of a strong $Li$ line is shown in Fig. \ref{EWLi-15806} (see also Table \ref{res}).

We cannot derive the $Li$ abundance since for these stars it is not trivial to derive spectroscopically a reliable estimate of the effective temperature. In fact, classically the temperatures can be derived from spectra in a wider spectral range: the spectra analyzed here do not show known spectral features suitable to derive the temperature. Besides, the effective temperatures derived from the observed colors are not reliable due to the peculiar position of the BWE stars in the color-magnitude diagram that can be altered by the scattering and/or obscuration effects related to the presence of the disk as well as by the accretion process at work. 
Since we cannot constrain the position in the CMD for the BWE stars, we cannot derive the $Li$ abundance.  

\begin{figure}[!t]
\centerline{\includegraphics[angle=0,width=8cm]{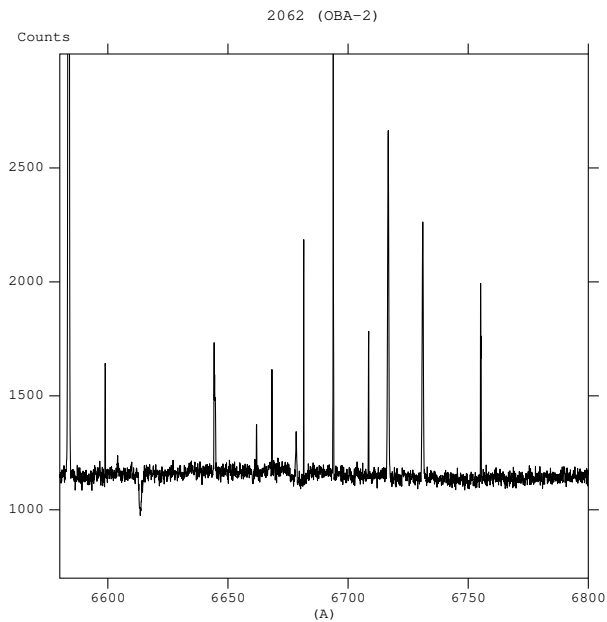}}
\caption{The spectrum of the BWE star $2062$ (excluding the spectral range near the $H\alpha$ line shown in Fig. \ref{Ha-profile-2062}) as observed during the OBA-2. This star is strongly influenced by veiling effects, as discussed in the text. In fact, the only one absorption line present in its spectrum is due to diffuse interstellar bands.}
\label{spettro-zoom-2062}
\end{figure}

We measured the $EW(Li)$ of the BWE stars, focusing on those stars showing a strong absorption line.
For those cases not showing an evident absorption $Li$ line, we investigated if this line can be affected by veiling effect (which causes a shallowing or a suppression of absorption lines), monitoring also other expected absorption lines in the region of the spectrum near the $Li$ line. 
If there is evidence of accretion features in the $H\alpha$ line profile, and if there are also only a few absorption lines observed in the spectrum, veiling effects could explain the reduction of the absorption lines.
As an example, the BWE star $2062$ suffers strong veiling effects. In fact, only one absorption line is visible in its spectrum (Fig. \ref{spettro-zoom-2062}) at $6613$ \AA, and it is due to diffuse interstellar bands (DIB). 
In contrast, spectra of unveiled BWE stars show many absorption lines, e.g. a strong $Li$ line (see Fig. \ref{EWLi-15806}).

\subsection{$H\alpha$ profiles}
\label{Ha}

We analyzed the $H\alpha$ profiles to investigate whether the disks associated with the confirmed BWE members (see last column in Table \ref{res} and discussion on membership in Sect. \ref{membership}) are active or inert disks.
In fact, for stars surrounded by active disks (i.e. with accretion and/or outflow processes still occurring), this emission line can be: a) symmetric, but showing broad wings, or b) asymmetric. In extreme cases a P Cygni profile
(i.e. a line with a deep blue absorption, under the continuum level, indicating a wind at work) or an inverse P Cygni profile (i.e. a deep red absorption, under the continuum level, indicating infall on the central star) can occur.

Following \citet{rpl96}, the $H\alpha$ line profile can be classified as: 1) type I, if the line is symmetric; 2) type II if the line is asymmetric, with a second emission peak whose intensity is greater than half of the main peak; 3) type III if the line is asymmetric with the second peak intensity lower that half of the main peak; 4) type IV if the line is asymmetric and there is an absorption feature below the continuum level (P Cygni profile). These features can be observed in the blue (B) or red (R) part of the main peak of the emission line. Therefore, a IV-R profile corresponds to the inverse P Cygni profile. Multiple profile (indicated by 'm') can also occur.
A possible physical explanation of these profile types is briefly discussed Sect. \ref{profili}.

From the analysis of the $H\alpha$ emission line profile and the comparison with the models, it is possible in general to estimate the accretion rate (\citealt{wb03}; \citealt{mhc03}). \citet{ntm04} have found that the measurement of the width of the line at $10\%$ of its peak, $H\alpha_{10\%}$, can be correlated to the accretion rate. However, this method cannot be used in the case of YSOs located near O stars, as in this case.
In fact, since their emission produces the formation of HII regions contributing to the $H\alpha$ emission, the line can contain a nebular contribution, concentrated in the region near the peak, which is also spatially variable and that cannot be easily measured independently.
Therefore, a good subtraction of the nebular contribution to the emission near the $H\alpha$ line cannot be achieved.
This prevent us from measuring the $EW(H\alpha)$ and the $H\alpha_{10\%}$ of the star excluding the nebular component.

However, the contribution to the emission line due to the nebula is narrow, while the CTTSs are expected to show a broad profile, since the $H\alpha$ line can also show the motion of the surrounding gas. 
Therefore, we measured the full width zero intensity ($FWZI$) of the $H\alpha$ line and we investigated its profile to select stars showing accretion or outflow signatures.
The stars with spectra showing narrow $H\alpha$ line (likely of nebular origin) are classified as candidate WTTSs, with an inert disk, since all the stars in the sample analyzed here are BWE, i.e with IR excesses by definition.

We analyzed the $50$ \AA~ range centered on the $H\alpha$ emission line of the  spectra. We normalized the spectra to the continuum level obtained by considering a second order Legendre function and varying the residual rejection limit in the IRAF task CONTINUUM. We performed the $FWZI$ measurements for each spectrum in all the different Configuration IDs (see Table \ref{obs-log}) to investigate also the emission variability, discussed in Sect. \ref{Ha variability}.

We considered the original spectra of the BWE stars without performing the sky subtraction. However, we have taken into account the nebular contribution on the $H\alpha$ emission. In fact, we evaluated the $\lambda_{min}$ and $\lambda_{max}$ wavelength range within which the nebular emission dominates the $H\alpha$ emission line. 
These values are quite consistent for all the OBs analyzed, ranging between $6561 - 6565$~\AA~with an average width $< 3$~\AA.

\begin{figure*}[!t]
\centerline{\includegraphics[angle=0,width=7cm]{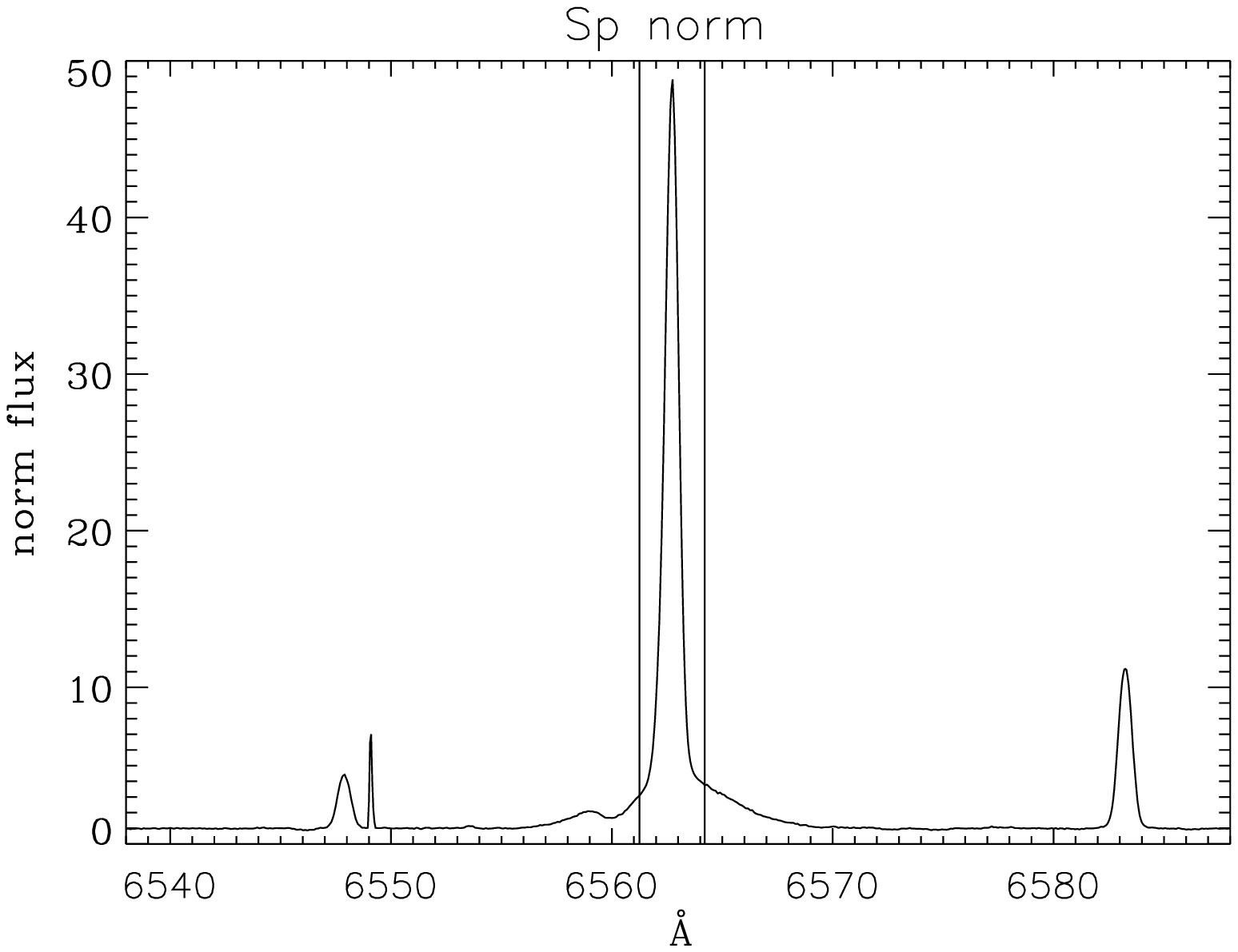}\hspace{-0.5cm}\includegraphics[angle=0,width=7cm]{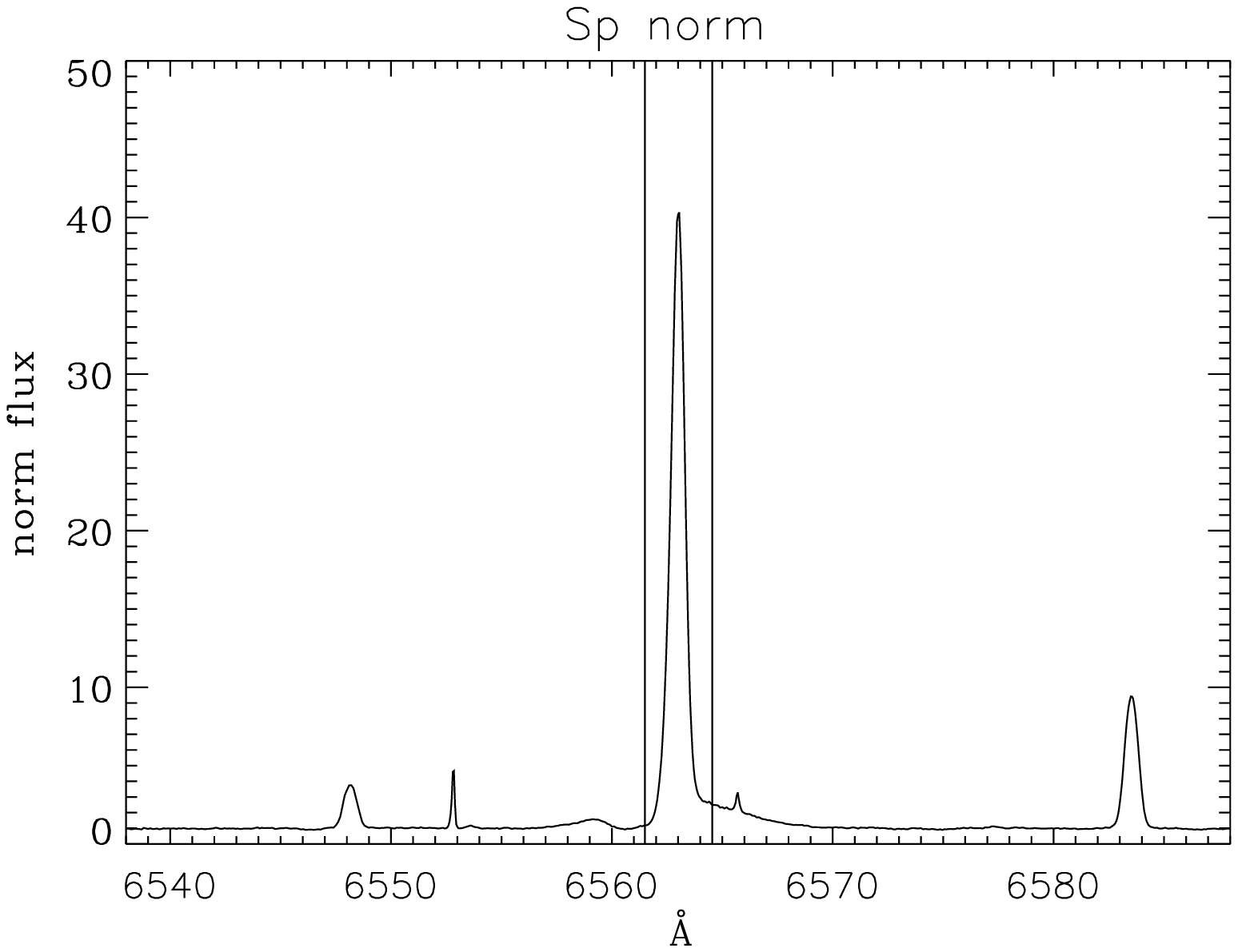}\hspace{-0.5cm}\includegraphics[angle=0,width=7cm]{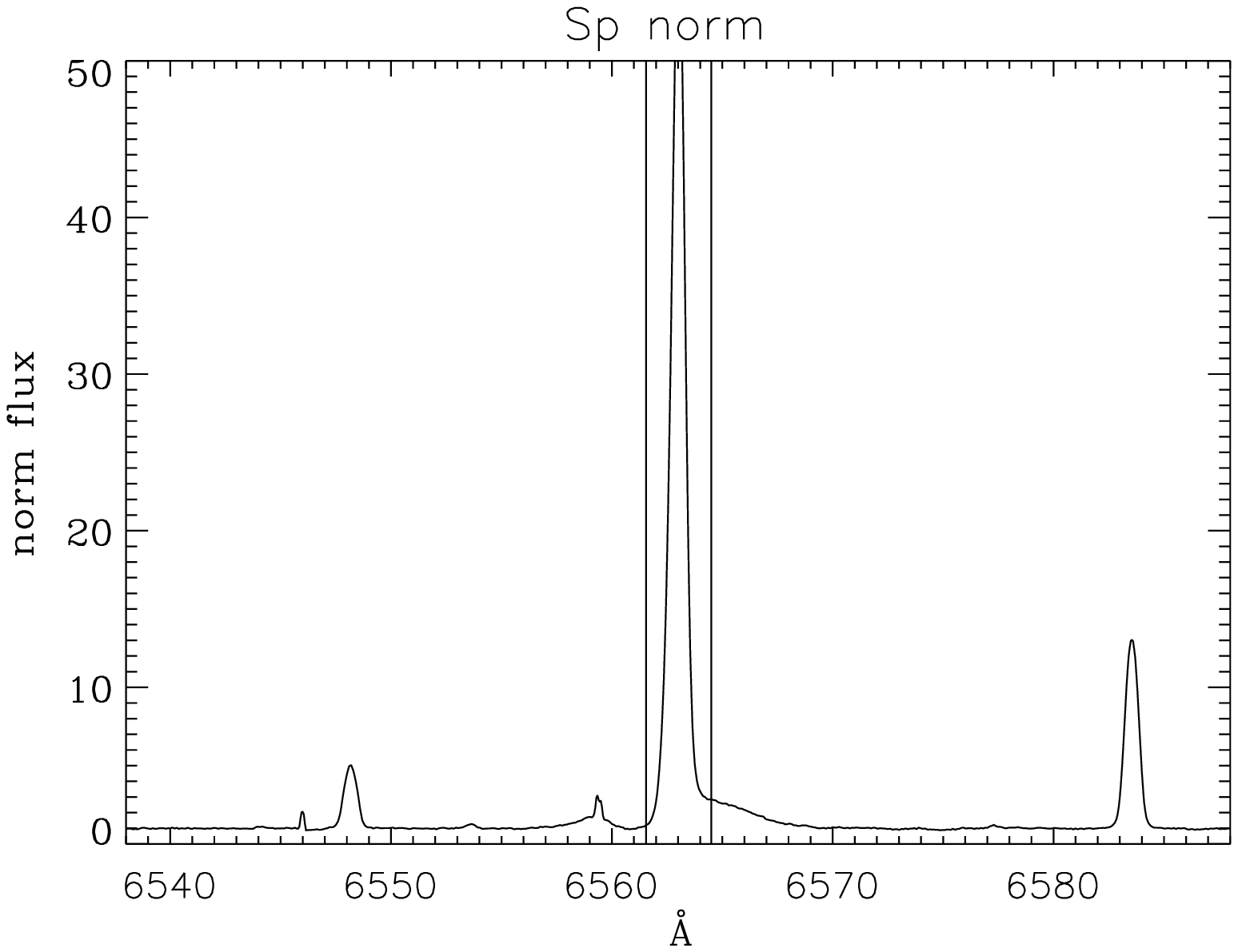}}
\caption{$H\alpha$ profiles of the $15584$ BWE star for all the OBs. For this object, even if we cannot distinguish the central profile, the $H\alpha$ profile clearly suggests accretion/outflow activity.}
\label{Ha-profile-15584}
\end{figure*}

\begin{figure*}[!t]
\centerline{\includegraphics[angle=0,width=7cm]{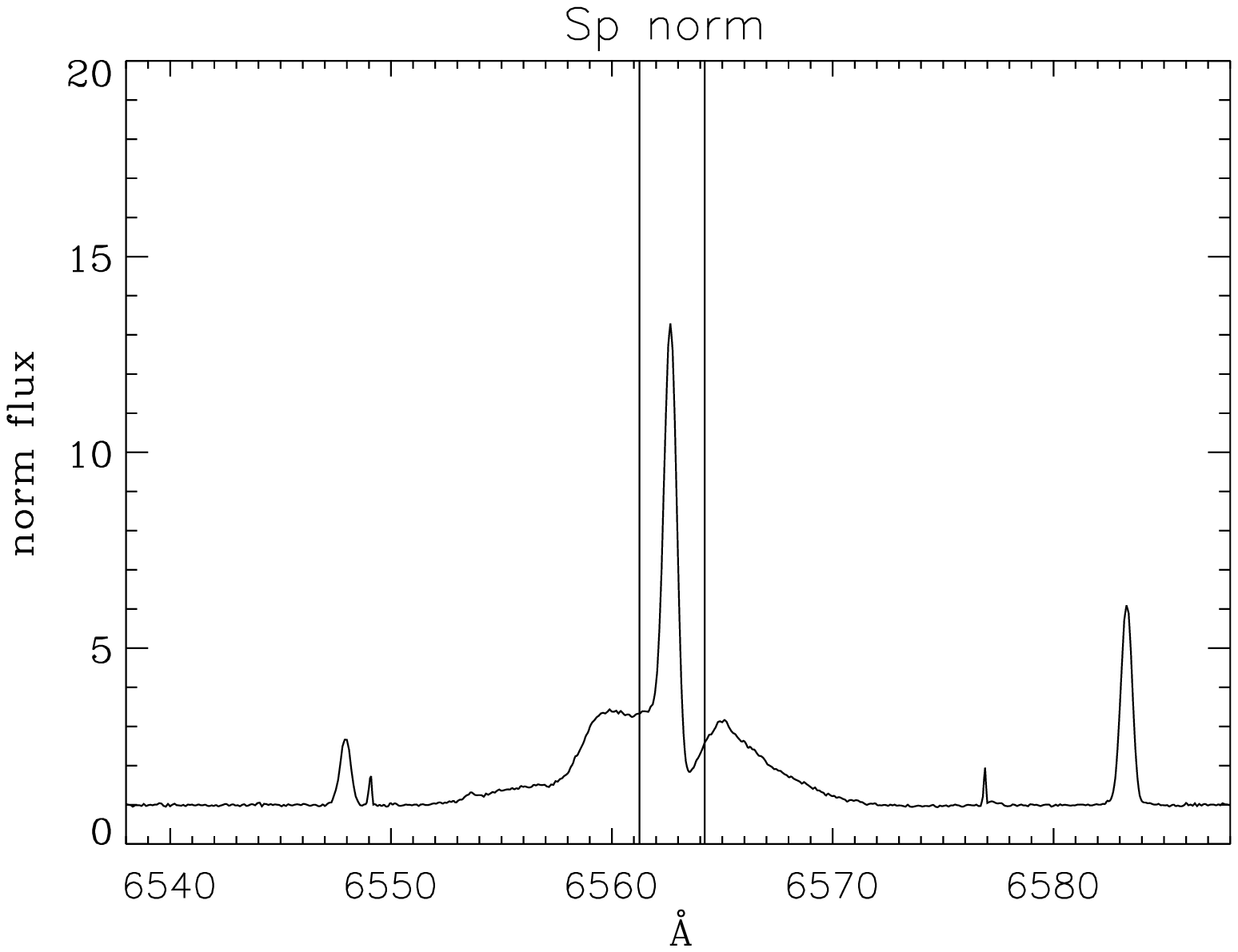}\hspace{-0.5cm}\includegraphics[angle=0,width=7cm]{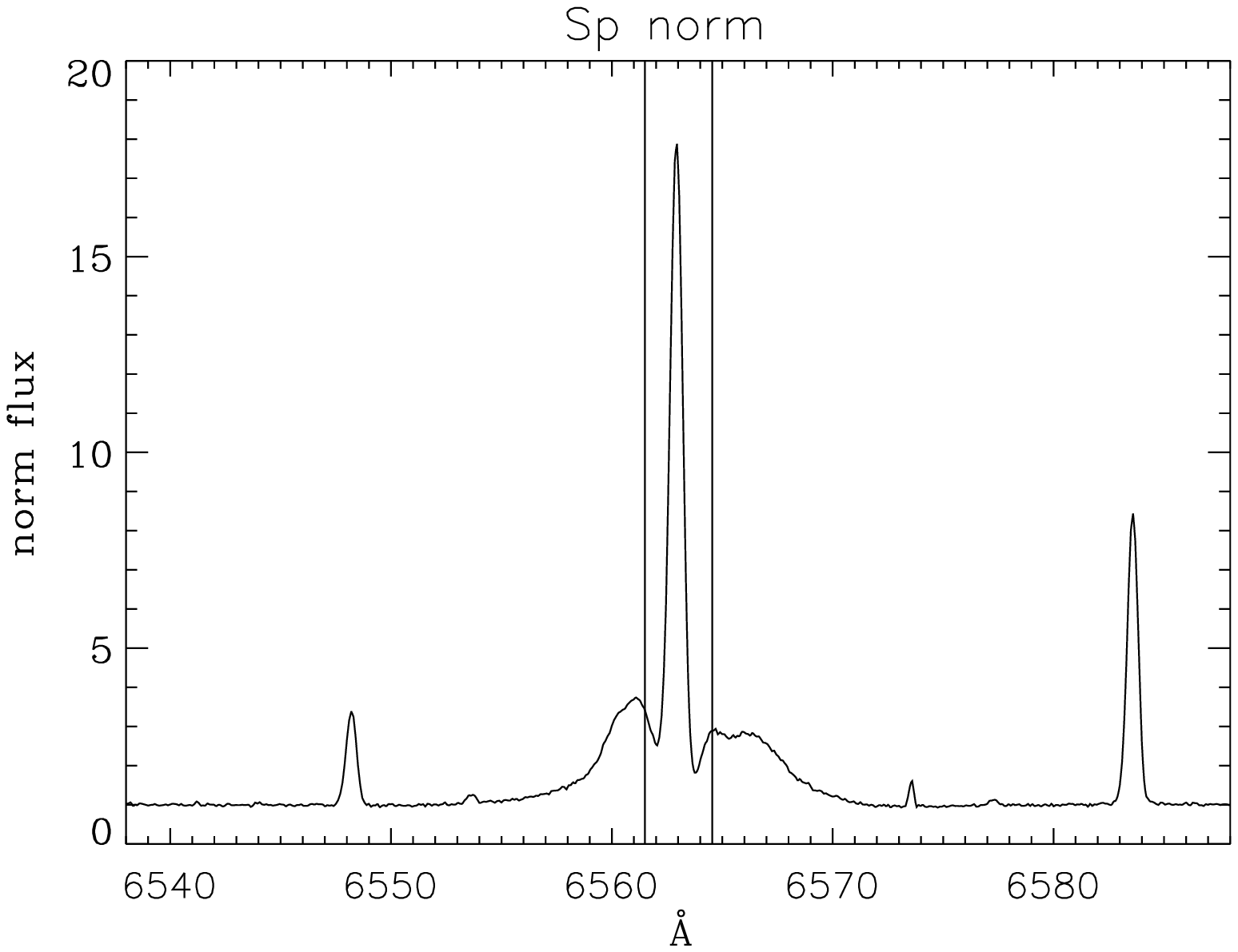}\hspace{-0.5cm}\includegraphics[angle=0,width=7cm]{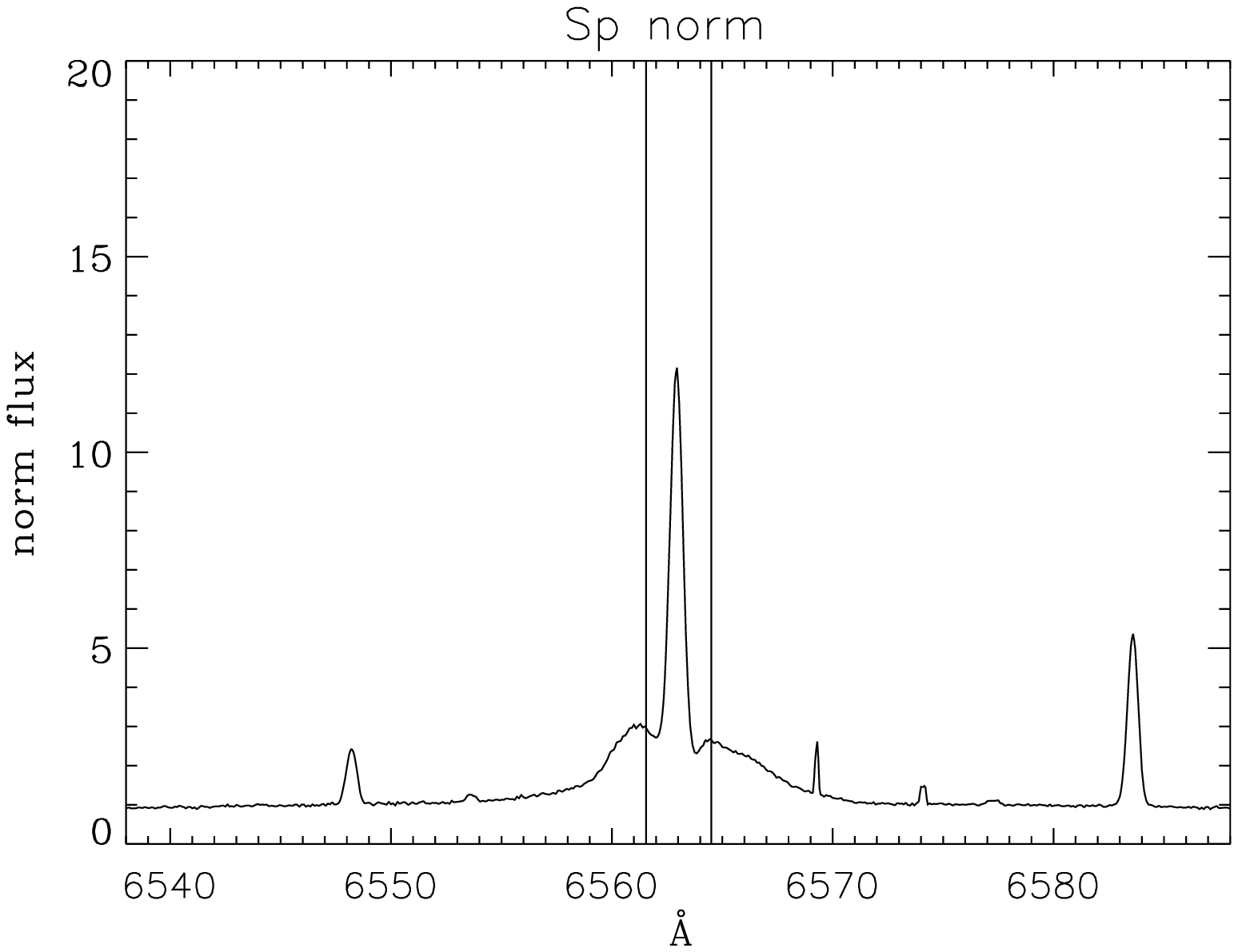}}
\centerline{\includegraphics[angle=0,width=7cm]{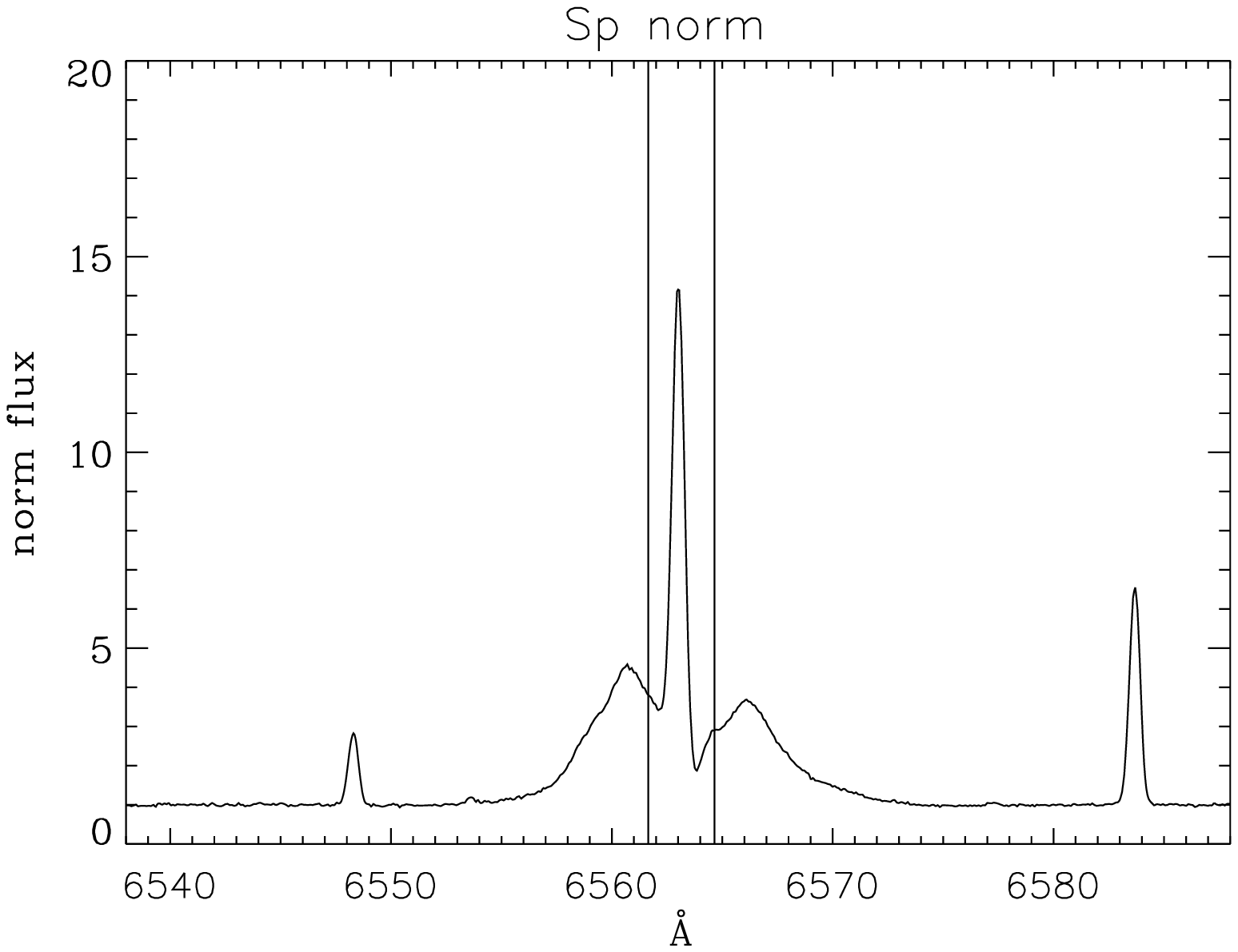}\hspace{-0.5cm}\includegraphics[angle=0,width=7cm]{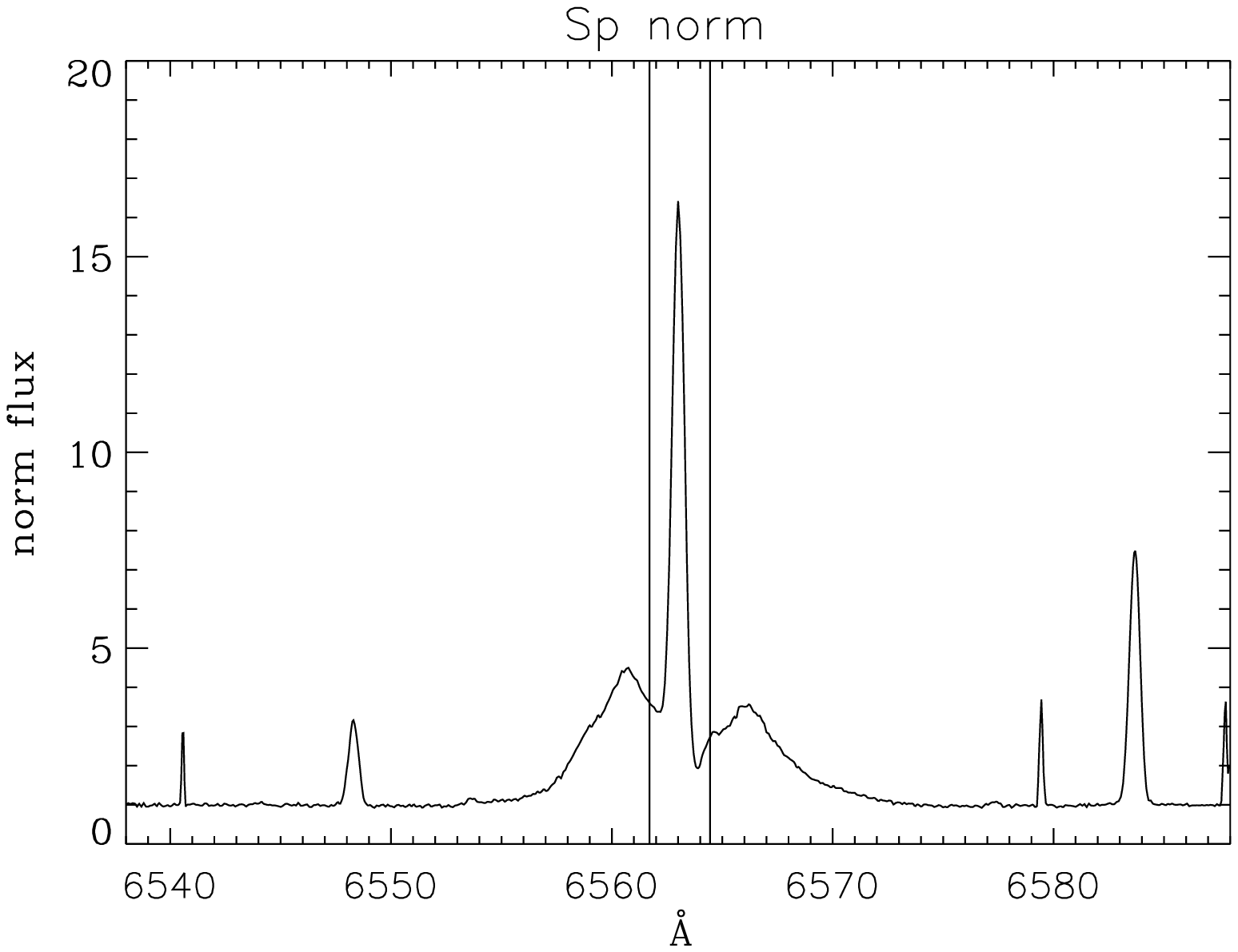}\hspace{-0.5cm}\includegraphics[angle=0,width=7cm]{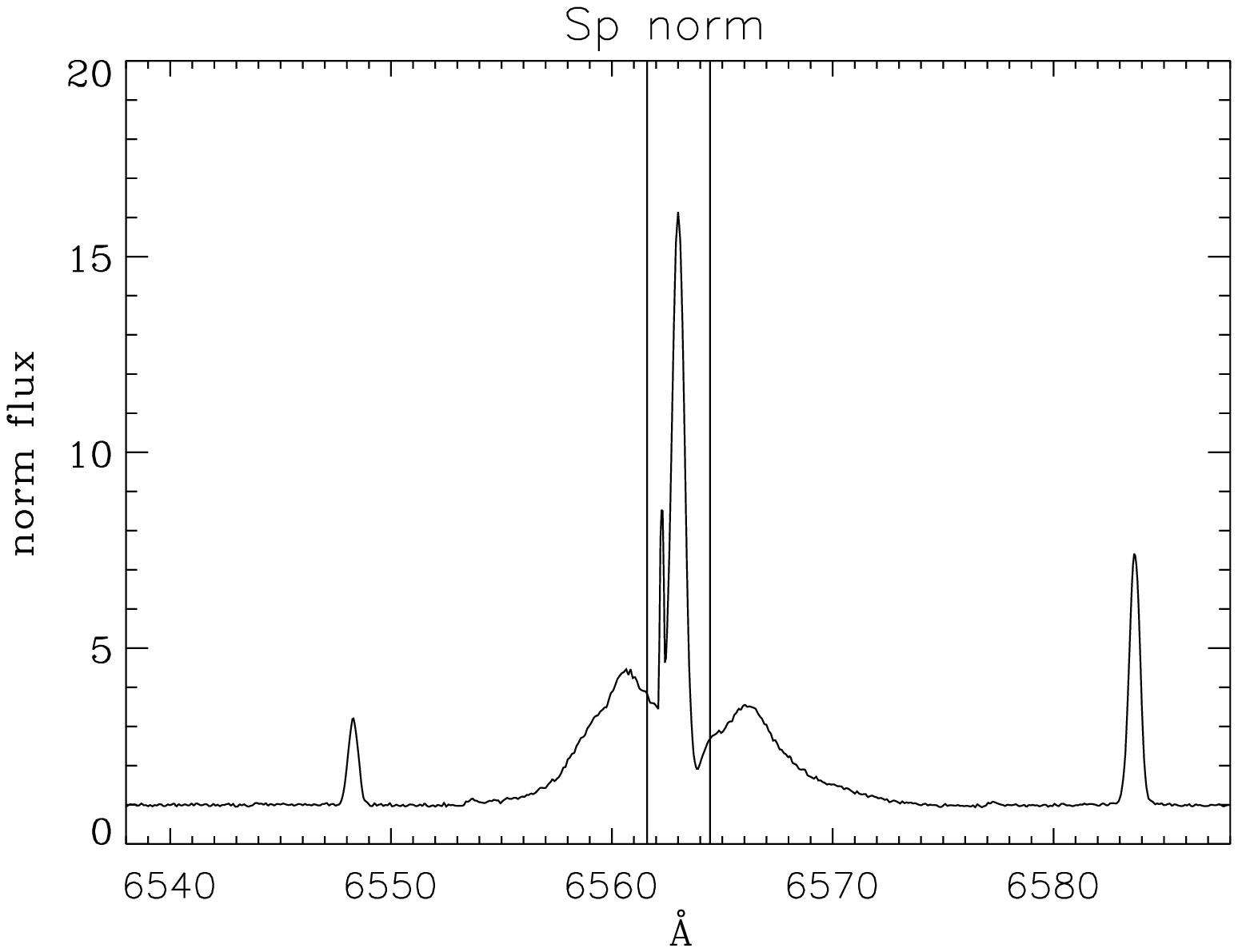}}
\caption{$H\alpha$ profiles of the $2062$ BWE star for all the OBs. For this object, even if we cannot distinguish the central profile, the $H\alpha$ profile clearly suggests accretion/outflow activity.}
\label{Ha-profile-2062}
\end{figure*}

In some cases, we can clearly recognize accretion/outflow activity in the $H\alpha$ line, even if we cannot distinguish the central profile dominated by the nebular emission (e.g. located between the two vertical lines superimposed on Fig. \ref{Ha-profile-15584}) and Fig. \ref{Ha-profile-2062}. 
For example, the BWE star $2062$ has a III-Bm or III-Rm profile\footnote{Note that we cannot discriminate between the cases of II and III profiles, both R or B, in general, if this profile lies within the nebular range}.

\begin{figure}[!t]
\centerline{\includegraphics[angle=0,width=10cm]{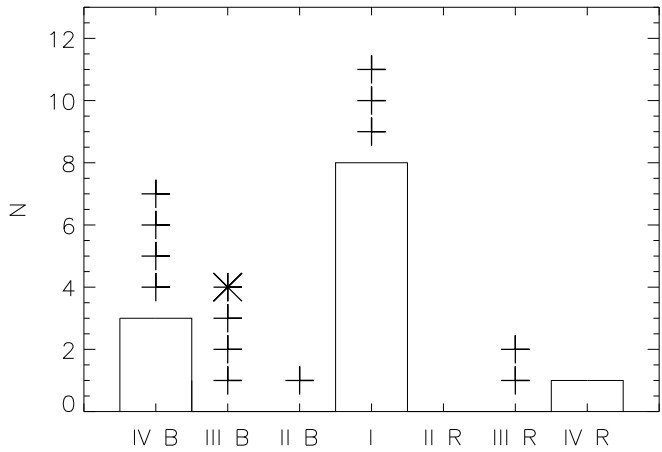}}
\caption{Histogram of the $H\alpha$ profile types. The crosses refer to the variable cases. The asterisks refer to the profile types different by the I type and not observable because of the nebular emission influence.}
\label{hist-Ha}
\end{figure}

Figure \ref{hist-Ha} shows the histogram of the profile types\footnote{See also Sect. \ref{Ha variability}, where we account for the variability in the profiles} observed in the BWE stars analyzed here. 
Even if the profile type I (corresponding to no evidence for disk activity) is the most frequent ($8$ stars plus $3$ variable, indicated by the crosses), in many cases the emission line profile indicates clearly accretion/outflow processes at work. In fact, the IV-B profile is evident in many spectra ($3 + 4$ variable), followed by the III-B type ($3$ variable $+ 1$ variable and within the nebular range, indicated by the asterisk). The III-R profile occurs twice (observed together with blue-shifted features); IV-R profile occurs once; II-B profile occurs once and it is variable.
In $2$ cases ($11751$ and $1320$) the spectra show a wide $H\alpha$ absorption line (with narrow emission due to the nebular contribution), evidence for these objects to be early stars, so no classification has been made for these stars.

\subsection{$H\alpha$ variability}
\label{Ha variability}

$H\alpha$ line profiles are expected to be variable since the infall, as well as the outflow, are not stationary processes. To investigate this variability, we compared all the spectra of each OB (different nights, see Table \ref{obs-log}) of the same star.

Table \ref{res} shows the profile type for each BWE star analyzed here. Some stars show more than one profile type (also indicated in the Table), suggesting a time variability of the $H\alpha$ profile.

Among the $20$ BWE stars discussed here\footnote{Excluding the two early-type stars, $11751$ and $1320$, see Table \ref{res}}, $12$ do not show variability in the $H\alpha$ emission line: $8$ show a type I profile, while in $4$ cases the line shows a P Cygni (or inverse P Cygni) profile.
 
An example of star showing variable emission line profile is $15584$, which varies from a III-Bm and a IV-Bm profile type.
In the BWE star $2062$ both intensity and position of the peaks vary, while the profile type is not variable (type III).
The variability of the $H\alpha$ emission line for the $15584$ and the $2062$ BWE stars is shown in Fig. \ref{Ha-profile-15584} and in Fig. \ref{Ha-profile-2062}.

\subsection{RV and vsini}
\label{RV and vsini}

In order to measure the radial velocity and projected rotational velocity for each BWE star, we used the FXCOR task of IRAF and performed a Fourier cross-correlation for each spectrum with respect to a template spectrum (following \citealt{td79}). The template star has been chosen among the stars without IR excess (not a BWE star), a ``bona fide" photometric member, with X-ray emission, with $V-I$ within the range of the BWE stars ($V-I = 1.2 - 2.2$ mag, being $(V-I)_{1660} = 2.02$ mag), and with a strong $Li$ line. 
The position of the peak of the cross-correlation function gives a measure of the radial velocity relative to the template (by selecting ranges without emission lines in the spectra). 

\begin{figure}[!t]
\centerline{\includegraphics[angle=0,width=8cm]{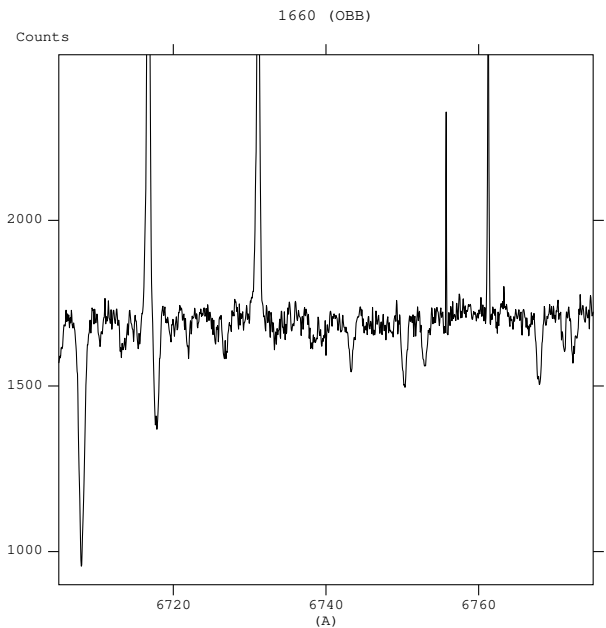}}
\caption{Spectrum near the $Li$ line (the strongest absorption line in this enlargement) of the template used to compute the $RV$ and $vsin(i)$ of all the BWE stars observed with FLAMES.}
\label{template}
\end{figure}

We have chosen the object $1660$ (see Fig. \ref{template}) as the template to derive the $RV$ and $vsin(i)$. 
The star $1660$ has been observed in all the OBs (and shows features of accretion/ejection in the $H\alpha$ profile but an adequate number of absorption lines to be used as a template). We compared the $RV$ values derived using the star $1660$ as template with those derived using another possible template, the star $16250$. 
We verified that the two templates provide consistent results. However, the template $1660$ ($V = 16.103$ mag, spectrum OBB-23) is the best choice, since it allows us to obtain smaller associated errors.

The cross-correlation function (CCF) can also appear to be broad due to rotation effect. Therefore its width is a measure of $vsin(i)$ of the star, if the template shows the minimum broadening. 
We derived the projected rotational velocities for our sample by using the CCF widths and interpolating these values with the relation between the width of the cross-correlation function and the rotational velocities $vsin(i)$ derived in \citet{pdm07} (by using synthetic spectra with spectral resolution very similar to that of our spectra).
The error associated with the $vsin(i)$ (following \citealt{rhm01}) is $\pm vsin(i)/(1 + r)$, where $r$ is the \citet{td79} parameter, which is a measure of the S/N of the peak of the CCF.  
The lower limit for the $vsin(i)$ measure is $17$ km/s, as this is the instrumental resolution.

For all the BWE stars analyzed here, we measured the $RV$ and $vsin(i)$ of each single observation (each OB), in order to search also for possible variability in the timescales involved in our observations (almost one month). Table \ref{res} shows our results. For those cases with spectra without enough absorption lines we cannot derive reliable values. For these objects no value is indicated in Table \ref{res}. 

The $vsin(i)$ for most BWE stars are $20 - 40$ km/s, a result which is consistent with the nature of T Tauri stars.

In addition we have found one case (star $4112$) in which the cross-correlation function is not symmetric and can be fitted with two Gaussians. This is an indication that this object is a double-lined spectroscopic binary (SB2).

In Table \ref{res}, the objects showing variability in their $RV$ are highlighted in bold.
A variable $RV$ in different OBs (different nights) suggests that the star can be a binary system and this is the case for $8$ objects.

\section{Discussion}
\label{Discussion}

\subsection{Membership}
\label{membership}

The criteria requested for a BWE star to be a candidate member of the NGC $6611$ cluster are: 1) the $EW(Li)$ is greater than $100$ m\AA; 2) the $H\alpha$ line profile indicates activity related to the accretion or outflow processes. 
A prominent absorption $Li$ line demonstrates the PMS nature of the object and the evidence of asymmetries and absorption feature in the $H\alpha$ emission line is a strong signature of the presence of an active disk surrounding the star.
We will also refer to previous detection of the stars in the X-ray band.
As for the $RV$ analysis, a value within $3\sigma$ with respect to the $RV$ of the template chosen is a criterion which is not strong enough in this case to discriminate between members and non members of the cluster because there could be a strong contamination which prevent us from using this criterion.

\begin{figure}[!t]
\centerline{\includegraphics[angle=-90,width=8cm]{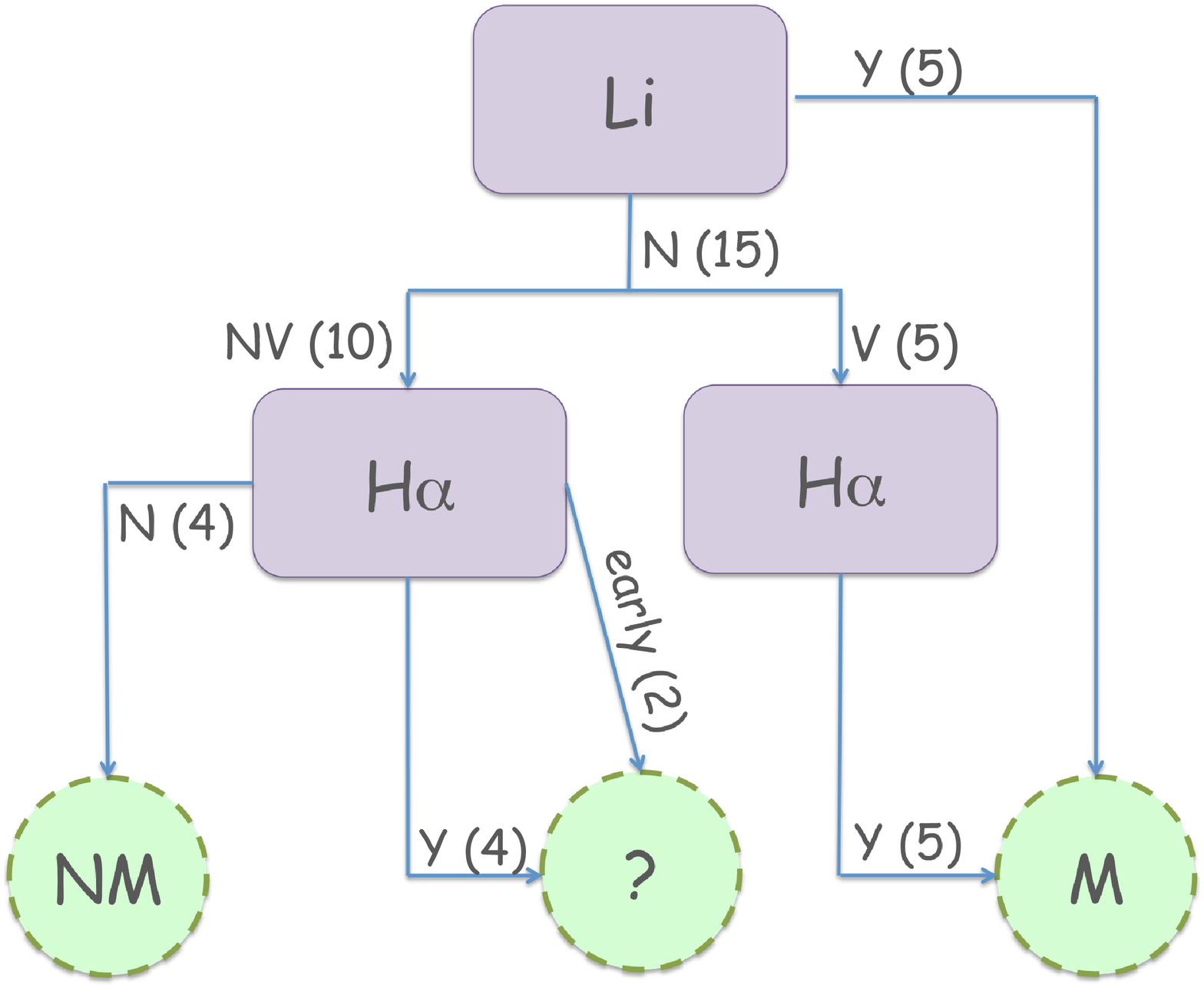}}
\caption{Membership criteria used to confirm the BWE stars as members or non-members: Y = yes; N = no; V = veiling; NV = no veiling; M = member; NM = non-member; ? = to be further investigated. In brackets the total number of objects for each case.}
\label{schema}
\end{figure}

To summarize the final membership of the BWE stars discussed here, we will refer to the flow chart shown in Fig. \ref{schema}.
First of all, we check for the evidence of the $Li$ absorption line in the spectra and confirm the BWE star as a cluster member if the $EW(Li)$ value is above the chosen threshold (``Y" in Table \ref{res} and in Fig. \ref{schema}, $5$ stars). 
This is a strong criterion (without considering other criteria to support the membership) because: 1) the IR excesses detected in the BWE stars suggest the presence of a circumstellar disk and 2) a symmetric $H\alpha$ emission line (type I profile) cannot exclude that the star is young.
In fact, if the $EW(Li)$ values derived are consistent with that object being a member of the cluster, but the $H\alpha$ emission line profile does not show features related to the accretion or outflow process, we can conclude that the object under inspection is a member of the cluster with an inert disk.
Note that if the $RV$ value of the object is not within $3\sigma$ with respect to the $RV$ of the template, but the $EW(Li)$ is above the threshold and there is also the detection of this object in X-rays (and this is true in $4$ cases over the total of $5$ BWE stars with evident $Li$), the star could be a binary member with the spectrum of just one component visible. Therefore, we do not reject this object to be a member of the cluster.

If the $EW(Li)$ measured from the spectra is below the chosen threshold (``N" in Table \ref{res} and in Fig. \ref{schema}, $15$ cases), we distinguish two different cases: the $Li$ line is not present possibly due to veiling (``V" in Fig. \ref{schema}, $5$ objects), or there is no evidence of veiling in the spectra (``NV" in Fig. \ref{schema}, $10$ objects). Then we check the $H\alpha$ emission line profile of the star.
If the $Li$ line is not evident possibly due to veiling effects on the spectra and the $H\alpha$ emission line profile is broad or asymmetric (``Y" in Table \ref{res} and in Fig. \ref{schema}, $5$ stars), then we consider this star as a member of the cluster. In fact, the presence of asymmetric features in this line is consistent with the presence of a circumstellar disk which can cause strong veiling effects due to accretion phenomena.  
If the $Li$ line is not strong and there is no evidence for veiling, then: a)  if there is not evidence of features related to accretion/outflow activity (``N" in Fig. \ref{schema}, $4$ cases), then this object is considered a non member; b) if the $H\alpha$ emission line profile is broad or asymmetric (``Y" in Fig. \ref{schema}), the object should be further investigated (``?" in Table \ref{res} and in Fig. \ref{schema}, $4$ cases); c) if the star is early-type ($2$ stars), in this case the $Li$ line is not expected to be strong, but the object should be further investigated (``?" in Table \ref{res} and in Fig. \ref{schema}, $2$ cases).

Following the considerations stated above, we can conclude that half of the sample analyzed here ($10/20$ objects) are confirmed members from our spectroscopic analysis of the BWE stars. 
In particular, among these $10$ BWE members: in $5$ cases there is evidence of a strong $Li$ absorption line, and four out of these five BWE stars with a strong $Li$ line also have been detected in X-rays ($15805$, $15806$, $13310$, and $15584$). The other BWE stars do not show X-ray detection ($11598$ was not in the field of view of the X-ray observations); in $5$ cases the $H\alpha$ line criterion is fulfilled, while the $Li$ line could be suppressed due to veiling effects. 

In $6$ cases we cannot discriminate between member or non member (``?" in Table \ref{res} and in Fig. \ref{schema}), since: in $4$ cases there is not evidence of a strong $Li$ line nor of veiling effects even if there are suggestions of accretion/outflow processes in the $H\alpha$ line; in $2$ cases the $H\alpha$ line analysis suggests these objects to be early-type stars (so we expect the $Li$ line not to be present in these stars).

Four cases are not confirmed as members of the NGC $6611$ cluster, since in these objects both criteria are not fulfilled.

\subsection{Binarity}
\label{Binarity}

As explained in Sect. \ref{membership}, we do not consider the $RV$ value consistent with the template as a strong membership criterion. However, we use the variability of $RV$ to check the binarity of the BWE stars. From our analysis we derive that $8$ BWE stars among the total of $20$ can be binary systems. In particular, the star $4112$ is possibly a SB2 as its cross correlation function has a double peaked profile. Among these $8$ binary stars: $3$ are confirmed members of NGC $6611$; for $3$ cases we cannot discriminate between member or non member; $2$ objects are not members of the cluster.

\subsection{Accretion properties: the physical origin of the $H\alpha$ profiles}
\label{profili}

In order to study the accretion properties of the BWE stars, we focused
on the $H\alpha$ line profiles observed with FLAMES. In general, the $H\alpha$ line profiles of young stars consist of a symmetrical emission line with one or more absorbing components superimposed, as anticipated in Sect. \ref{Ha}. Here we investigate the physical origin of the $H\alpha$ profile types detected in the BWE stars, following \citet{rpl96}.
The $H\alpha$ line consists of several components to which different regions (characterized by different dynamical and physical conditions) contribute.
Magnetically driven winds can explain the central and the blue-shifted absorption. On the other hand, the free fall of material from the disk onto the star (leaded by a magnetic field) causes the infalling gas to reach velocity up to several hundred km/s, decelerated in strong shocks which can explain the asymmetry of the $H\alpha$ line due to a red-shifted absorption component. 
Models of winds accelerated close to the star as blobs and then decelerated can reproduce the double peaks observed.
Therefore, type IV-B or P Cygni profiles can be explained by models of spherical winds, while type III-B profiles are reproduced by stochastic decelerating wind models.  
Infall material can explain type II and III-R profiles if different parameters are used in the model (see all the profile type first presented in Sect. \ref{Ha}).

\citet{khs06} interpret the profile types defined in \citet{rpl96} also in term of inclination and accretion/outflow activity. In particular, the type II-B can be reproduced by models with medium-to-high inclinations and are due to fast wind acceleration. 
The difference between Type II-R profile and type II-B is interpreted as a geometrical effect corresponding to different viewing angles.
Type III-B profiles are reproduced by models with moderate inclination and due to fast wind acceleration, while type III-R are associated with high inclination (the disk obscures the central emission).
The classical P Cygni profile (type IV-B) where the component due to the absorption in the blue has a sufficient velocity to be beyond the emission line wing, can be explained with a bipolar flow observed along its axis, while the inverse P Cygni profile (type IV-R) is related to the lowest mass-accretion rates at high inclinations.

In the context of the BWE stars detected in NGC $6611$ and analyzed here, absorption features superimposed on the emission line has been detected in $10$ BWE stars among the total of $20$ discussed, and lead to the profile type II, III, or IV. In some cases also the type I profile can be asymmetric itself indicating activity of the surrounding disk, and this occurs in the BWE star $1336$ (see also Table \ref{res}).
The absorption can be due to material with a large range of velocity. In fact, the dips observed can be narrow (as in the case of $11532$, OBA) or broad (e.g. $2062$). The multiple profiles ('m') are characterized by several dips or absorption features (as in the BWE star $15584$).

\subsection{BWE stars: old members?}
\label{old}

There are several interpretations of the nature of the BWE stars (see Sect. \ref{Introduction}).
A qualitative analysis of the SEDs of this sample of stars suggests that obscuration or scattering due to the presence of the surrounding disk can explain the observed colors of some BWE stars. On the other hand,
these BWE stars could actually be cluster members significantly older than the mean cluster age (suggesting an age spread as in \citealt{prf05}), i.e. a first generation of cluster members yet surrounded by a circumstellar disk. In this case, their colors are purely photospheric, in fact there are other examples of CTTSs older than the typical lifetime of the disk (e.g. MP Mus, \citealt{amp07}).
Our data do not allow us to discard this hypothesis since the $EW$ of the $Li$ is compatible with that of stars where a partial  (less than a factor $10$) $Li$ depletion occurred. On the other hand, given the absence of molecular bands, our targets are certainly stars with spectral type earlier or equal to K stars for which a partial and slow $Li$ depletion is expected up to $10$ Myr. Nevertheless, the measured $EW$ of the $Li$ cannot be converted accurately in $Li$ abundances since the effective temperatures of these objects cannot be derived with available data and thus we are not able to constrain their ages by using the $Li$ depletion timescale inferred by stellar models.

\section{Summary and conclusions}
\label{Summary and conclusions}

In this paper, we have investigated the nature of the BWE stars. This is a class of candidate members to young clusters showing IR excesses, signature of the presence of a circumstellar disk, but with a photometric position inconsistent with the locus defined by the young cluster members. 
Here we have reported the spectroscopic analysis of twenty BWE stars observed with FLAMES to confirm their membership to the NGC $6611$ cluster and to understand the origin of their position in the CMD.

We have analyzed the $H\alpha$ profiles and derived that in $10/20$ stars the emission line shows absorption features or asymmetry, indicative of the presence of an active disk.  
In summary, half of the analyzed BWE stars presents accretion and outflow with various characteristics.

From the study of the $Li$ line, we derived that in $5$ cases there is evidence of a strong absorption $Li$ line, suggesting that these are young objects.

The study of the $RV$ allowed us to infer the binarity of $8$ BWE stars, $3$ of which confirmed as members of the cluster.

We have taken advantage of the spectroscopic results derived from the analysis of the $Li$ absorption line and the $H\alpha$ profiles to confirm the membership of the BWE stars to NGC $6611$.
Ten ($50 \%$ of the total sample) BWE stars have been confirmed as members of the cluster. Among these, we interpret the $4$ stars showing $H\alpha$ type I profiles ($12168$, $13310$, $15805$, and $15806$) as WTTSs with an inert disk. The other $6$ confirmed members have IV-B profiles or variable III m or IV-B profile types, therefore these BWE stars can be classified as CTTSs with active accretion disk.
The $H\alpha$ profiles observed in the $6$ cases where the membership is uncertain suggest: in $2$ cases that the objects are early-type stars; in $4$ cases the presence of accretion or outflow activity (with absorption feature in the blue and red part of the wings). Therefore, the latter could be classified as CTTSs with active accretion disks, if confirmed as members of the cluster. 

The origin of the peculiar position of the BWE stars in the color-magnitude diagram can be explained as due to the scattering and/or obscuration effects related to the presence of the disk as well as by the accretion process at work. On the other hand, we cannot exclude the hypothesis of an older population of the cluster (De Marchi et al. in preparation), but we can conclude unambiguously that about half of the BWE stars are members of the cluster.
In conclusion, in fact, a robust result of our work is that the spectroscopic analysis (as in this case, using FLAMES data) of BWE stars unambiguously demonstrates that half ($10$ stars out of the total $20$ stars of the selected sample) of BWE stars has been confirmed as members of the cluster, with $6$ more objects that are possible members as well.

Several young clusters, apart from NGC $6611$ discussed here, host stars with blue optical colors consistent with field stars or old cluster members, as Orion Nebula Cluster (\citealt{ps00}), NGC $6530$ (\citealt{pdm05}), NGC $2264$ (\citealt{fms00}), NGC $1893$ (\citealt{psm11}). 
Our analysis suggests that not including the BWE stars in the membership of young clusters with high disk fraction results in non completed membership, and that the investigation of the spectroscopic properties of the BWE stars can be of general interest in the context of young clusters.

\begin{acknowledgements}

We would like to thank the referee for her/his useful comments.
This work was supported in part by Agenzia Spaziale Italiana under contract No. ASI-INAF (I/009/10/0).
L. P. and G. M. acknowledge the PRIN-INAF (PI. Desidera) for financial support.
M.G.G. is supported by the Chandra grant GO0-11040X.
We would like to thank Dr. D. Randazzo for her suggestions during the preparation of the manuscript.
\end{acknowledgements}

\bibliographystyle{aa}
\bibliography{references}

%\begin{longtab}
\begin{landscape}
\begin{longtable}{cccccccccccc}
\caption{\label{res} Results for the $20$ BWE stars discussed here, for all the OBs. ID of each BWE star; OB in which each BWE star has been observed (number of the spectrum in brackets); $H\alpha$ profile type (in brackets the profiles not visible because of the nebular influence, as explained in the text); $FWZI$ of this emission line ($-$ if the profile is within the nebular range); $EW(Li)$; $RV$ and $vsin(i)$ ($-$ for cases in which there are not enough lines that can be used in FXCOR). In bold are highlighted the objects showing a variable $RV$. Detection in X-rays ($-$ if outside of the field of view); total membership using the criteria: $RV$, $Li > 100$ m\AA, and $H\alpha$ emission line (N (V?): veiling?; Y: YES; N: NO; SKY: emission line within the $\lambda$ range effected by the nebular emission, B?: binary?; SB2: binary system with the two components resolved); the values that are the same for each of the OBs are shown in the first OB in which the star is detected; otherwise is indicated).}\\
\hline\hline
ID             & OB (Sp)    & $H\alpha$              & $FWZI$       & $EW(Li)$        & $RV$            & $v sin(i)$        & X & $RV$     &        & MEMBERSHIP &     \\
               &            & profile type           & (\AA)        & (m\AA)          & (km/s)          & (km/s)            &   &          & $Li$   & $H\alpha$  & TOT \\   
\hline
\endfirsthead
\caption{continued.}\\
\hline\hline
ID             & OB (Sp)    & $H\alpha$              & $FWZI$       & $EW(Li)$        & $RV$            & $v sin(i)$        & X & $RV$     &        & MEMBERSHIP &     \\
               &            & profile type           & (\AA)        & (m\AA)          & (km/s)          & (km/s)            &   &     & $Li$   & $H\alpha$  & TOT \\   
\hline
\endhead
\hline\hline
\endfoot
10151          & OBA-2 (62) & (I)                    & $-$          & $-$             & $-9.45\pm3.27$  & $17.7\pm1.7$      & N & Y    & N      & N          & N   \\  
10151          & OBB-2 (51) & (I)                    & $-$          & $-$             & $-10.91\pm2.91$ & $< 17$            &   &      &        &            &     \\ 
10151          & OBC-2 (51) & (I)                    & $-$          & $-$             & $-9.68\pm2.56$  & $< 17$            &   &      &        &            &     \\  
\hline

11447          & OBA (74)   & IV-B                   & $2.8-3.9$    & $-$             & $-51.48\pm3.43$	& $21.1\pm2.1$      & N & N    & N (V?) & Y	     & Y   \\
11447          & OBB (74)   & IV-B                   &              & $-$             & $-38.38\pm3.50$	& $19.4\pm2.0$		    &   &      &        &	     &	   \\
11447          & OBC (74)   & IV-B                   &              & $-$             & $-$             & $-$               &   &      &        &            &     \\
\hline

\textbf{11532} & OBA (75)   & (III-B)                & $-$          & $-$             & $10.69\pm2.46$  & $20.3\pm1.4$      & N & B?   & N      & Y SKY      & ?   \\
\textbf{11532} & OBB (76)   & (I)                    & $-$          & $-$             & $-$             & $-$               &   &      &        &            &     \\
\textbf{11532} & OBC (76)   & (I)                    & $-$          & $-$             & $26.21\pm2.61$  & $< 17$            &   &      &        &            &     \\
\hline

11598          & OBA (64)   & IV-B                   & $4.1-4.2$    & $-$             & $-71.22\pm3.77$ & $< 17$            & - & N    & N (V?) & Y          & Y   \\
11598          & OBB (71)   & IV-B                   &              & $-$             & $-58.63\pm5.15$ & $< 17$            & - &      &        &            &     \\
11598          & OBC (80)   & IV-B                   &              & $-$             & $-59.95\pm4.67$ & $< 17$            & - &      &        &            &     \\
\hline

11751          & OBA (88)   & early                  &              & $-$             & $-$             & $-$               & N & -    & N      & early      & ?   \\
11751          & OBA-2 (82) & early                  &              & $-$             & $-$             & $-$               &   &      &        &            &     \\
11751          & OBB (73)   & early                  &              & $-$             & $-$             & $-$               &   &      &        &            &     \\
\hline

\textbf{1211}  & OBA (51)   & (I)                    & $-$          & $-$             & $-25.20\pm2.11$ & $< 17$            & N & B?   & N      & N          & N   \\
\textbf{1211}  & OBA-2 (42) & (I)                    & $-$          & $-$             & $-6.88\pm1.97$	& $< 17$            &   &      &        &            &     \\
\textbf{1211}  & OBB (51)   & (I)                    & $-$          & $-$             & $-12.25\pm2.39$ & $< 17$            &   &      &        &            &     \\
\textbf{1211}  & OBB-2 (42) & (I)                    & $-$          & $-$             & $-8.29\pm7.35$	& $33.9\pm5.9$      &   &      &        &            &     \\
\textbf{1211}  & OBC (49)   & (I)                    & $-$          & $-$             & $-10.66\pm4.23$ & $< 17$            &   &      &        &            &     \\
\textbf{1211}  & OBC-2 (42) & (I)                    & $-$          & $-$             & $-7.01\pm6.72$	& $34.4\pm5.4$      &   &      &        &            &     \\
\hline

\textbf{12168} & OBA (52)   & (I)                    & $-$          & $185$           & $-33.92\pm2.29$ & $17.6\pm1.2$      & N & B?   & Y      & N          & Y   \\
\textbf{12168} & OBA-2 (72) & (I)                    & $-$          &                 & $-17.34\pm2.11$ & $21.5\pm1.3$      &   &      &        &            &     \\
\textbf{12168} & OBC (62)   & (I)                    & $-$          &                 & $-21.30\pm2.10$ & $< 17$            &   &      &        &            &     \\
\hline

1233           & OBA-2 (30) & (I)                    & $-$          & $-$             & $-20.37\pm2.30$ & $19.9\pm1.3$      & N & N    & N      & N          & N   \\
1233           & OBB-2 (40) & (I)                    & $-$          & $-$             & $-20.80\pm2.35$ & $22.3\pm1.5$      &   &      &        &            &     \\
1233           & OBC-2 (40) & (I)                    & $-$          & $-$             & $-21.04\pm2.57$ & $21.2\pm1.5$      &   &      &        &            &     \\
\hline

\textbf{12631} & OBA (63)   & (I)                    & $-$          & $-$             & $-20.85\pm2.02$ & $17.9\pm1.1$      & N & B?   & N      & N          & N   \\
\textbf{12631} & OBA-2 (76) & (I)                    & $-$          & $-$             & $-4.09\pm2.33$	& $21.0\pm1.4$      &	&      &        &            &     \\
\textbf{12631} & OBB (63)   & (I)                    & $-$          & $-$             & $-9.33\pm2.14$	& $< 17$            &   &      &        &            &     \\
\textbf{12631} & OBB-2 (73) & (I)                    & $-$          & $-$             & $-4.85\pm1.88$	& $< 17$            &   &      &        &            &     \\
\textbf{12631} & OBC (70)   & (I)                    & $-$          & $-$             & $-9.63\pm3.33$	& $< 17$            &   &      &        &            &     \\
\textbf{12631} & OBC-2 (73) & (I)                    & $-$          & $-$             & $-6.20\pm2.12$	& $< 17$            &   &      &        &            &     \\
\hline

1320           & OBA-2 (40) &  early                 &              & $-$             & $-10.06\pm5.87$ & $20.4\pm3.4$      & N & Y    & N      & early      & ?   \\
1320           & OBB-2 (33) &  early                 &              & $-$             & $-9.49\pm5.85$	& $17.6\pm3.0$      &   &      &        &            &     \\
1320           & OBC-2 (33) &  early                 &              & $-$             & $-9.57\pm5.47$	& $18.1\pm2.9$      &   &      &        &            &     \\
\hline

13310          & OBA (69)   & (I)                    & $-$          &                 & $-18.24\pm3.49$	& $37.2\pm2.9$      & Y & N    & Y      & N	     & Y  \\
13310          & OBB (64)   & (I)                    & $-$          &                 & $-6.86\pm1.75$	& $33.3\pm1.4$      &   &      &        &	     &	 \\
13310          & OBC (60)   & (I)                    & $-$          & $284$           & $-$             & $-$               &   &      &        &            &    \\
\hline

\textbf{1336}  & OBA (41)   & IV-B                   & $2.7-3.4$    & $-$             & $-32.84\pm2.28$ & $17.7\pm1.2$      & N & B?   & N      & Y          & ?  \\
\textbf{1336}  & OBA-2 (34) & IV-B                   &              & $-$             & $-15.75\pm2.38$ & $17.1\pm1.2$	    &   &      &        &            &    \\
\textbf{1336}  & OBB (42)   & IV-B                   &              & $-$             & $-21.01\pm2.56$ & $20.0\pm1.5$	    &   &      &        &            &    \\
\textbf{1336}  & OBB-2 (38) & IV-B                   &              & $-$             & $-15.55\pm2.37$ & $19.6\pm1.3$	    &   &      &        &            &    \\
\textbf{1336}  & OBC (33)   & (I asymm.)             &              & $-$             & $-19.45\pm3.56$	& $17.8\pm1.9$      &   &      &        &            &    \\
\textbf{1336}  & OBC-2 (38) & IV-B                   &              & $-$             & $-15.71\pm2.29$ & $< 17$	    &   &      &        &            &    \\
\hline

13407          & OBB-2 (98) & (I)                    & $3.6$        & $-$             & $-44.17\pm2.53$ & $20.3\pm1.5$      & N & N    & N (V?) & Y          & Y  \\
13407          & OBC (65)   & IV-B                   &              & $-$             & $-48.29\pm2.39$ & $22.3\pm1.5$      &   &      &        &            &    \\
13407          & OBC-2 (98) & (I)                    &              & $-$             & $-43.97\pm2.80$ & $22.2\pm1.7$      &   &      &        &            &    \\
\hline

1455           & OBA-2 (41) & IV-Bm                  & $4.1-4.4$    & $-$             & $-$             & $-$               & N & N    & N (V?) & Y          & Y  \\
1455           & OBB-2 (36) & IV-Bm                  &              & $-$             & $-52.09\pm3.52$	& $23.2\pm2.3$	    &   &      &        &	     &	 \\
1455           & OBC-2 (36) & IV-Bm                  &              & $-$             & $-51.31\pm3.47$	& $24.7\pm2.3$		    &   &      &        &	     &	 \\
\hline

\textbf{15584} & OBA (86)   & III-Bm                 & $11.6-14.6$  & $324$ (OBA,OBC) & $-10.88\pm2.06$ & $40.0\pm1.8$      & Y & B?   & Y      & Y          & Y  \\
\textbf{15584} & OBB (86)   & IV-Bm + III-R          &              &                 & $0.32\pm3.33$	& $38.7\pm2.8$      &   &      &        &            &    \\
\textbf{15584} & OBC (92)   & IV-Bm                  &              &                 & $0.94\pm2.12$	& $41.0\pm1.8$      &   &      &        &            &    \\
\hline

15805          & OBA-2 (90) & (I)                    & $-$          & $392$ (OBA-2,OBC-2) & $5.64\pm2.35$ & $40.7\pm2.0$    & Y & Y    & Y      & N          & Y  \\
15805          & OBB-2 (94) & (I)                    & $-$          &                 & $8.17\pm3.04$	& $43.1\pm2.7$      &   &      &        &            &    \\
15805          & OBC-2 (94) & (I)                    & $-$          &                 & $5.55\pm2.79$	& $42.5\pm2.4$      &   &      &        &            &    \\
\hline

\textbf{15806} & OBA (96)   & (I)                    & $-$          & $448$           & $-18.65\pm3.14$	& $34.7\pm2.5$      & Y & B?   & Y      & N          & Y  \\
\textbf{15806} & OBB (85)   & (I)                    & $-$          &                 & $-7.18\pm2.7$	& $31.1\pm2.1$      &   &      &        &            &    \\
\textbf{15806} & OBC (86)   & (I)                    & $-$          &                 & $-7.39\pm2.16$	& $33.7\pm1.7$      &   &      &        &            &    \\
\hline

2062           & OBA (40)   & III-B or Rm            & $17.9-21.3$  & $-$             & $-$             & $-$               & N & -    & N (V)  & Y          & Y  \\
2062           & OBA-2 (29) & III-B or Rm            &              & $-$             & $-$             & $-$               &   &      &        &            &    \\
2062           & OBB (40)   & III-B or Rm            &              & $-$             & $-$             & $-$               &   &      &        &            &    \\
2062           & OBB-2 (24) & III-B or Rm            &              & $-$             & $-$             & $-$               &   &      &        &            &    \\
2062           & OBC (37)   & III-B or Rm            &              & $-$             & $-$             & $-$               &   &      &        &            &    \\
2062           & OBC-2 (24) & III-B or Rm            &              & $-$             & $-$             & $-$               &   &      &        &            &    \\
\hline

2594           & OBA-2 (3)  & IV-Bm                  & $4.3-11.4$   & $-$             & $-48.75\pm3.13$ & $< 17$            & N & N    & N      & Y          & ?  \\
2594           & OBB-2 (4)  & III-Bm                 &              & $-$             & $-47.82\pm5.61$ & $25.3\pm3.8$      &   &      &        &            &    \\
2594           & OBC-2 (4)  & II-Bm                  &              & $-$             & $-47.59\pm5.77$ & $23.8\pm3.8$      &   &      &        &            &    \\
\hline
 
\textbf{4112}  & OBA (7)    & IV-Rm                  & $3.3-4.7$    & $-$             & $36.98\pm2.68$	& $26.9\pm1.9$      & N & SB2  & N      & Y          & ?  \\
\textbf{4112}  & OBB (7)    & IV-Rm                  &              & $-$             & $50.23\pm2.90$	& $23.1\pm1.8$      &   &      &        &            &    \\
\textbf{4112}  & OBC (5)    & IV-Rm                  &              & $-$             & $50.62\pm3.50$  & $20.6\pm2.1$      &   &      &        &            &    \\
\end{longtable}
\end{landscape}
%\end{longtab}

\end{document}